\documentclass[aps,prx,twocolumn,superscriptaddress,floatfix,longbibliography]{revtex4-2}
\usepackage{amsfonts,amsmath,amsthm,amssymb,amsopn,graphicx}
\usepackage{dsfont,bbm,xcolor}
\usepackage{enumitem}
\usepackage{hyperref}
\usepackage{breqn}
\usepackage{siunitx}
\usepackage{color}
\usepackage[]{breqn}
\usepackage{mathtools}
\usepackage{bigints}
\usepackage[caption=false]{subfig}

\makeatletter
\let\cat@comma@active\@empty
\makeatother

\def\bra#1{\mathinner{\langle{#1}|}}
\def\ket#1{\mathinner{|{#1}\rangle}}
\def\abs#1{\left | #1 \right |}
\newcommand{\braket}[2]{\langle #1 \vert #2 \rangle}

\def\bra#1{\mathinner{\langle{#1}|}}
\def\ket#1{\mathinner{|{#1}\rangle}}

\newcommand{\norm}[1]{\left\lVert#1\right\rVert}

\def\Im{{\,{\rm Im}\,}}

\newcommand{\intbz}{\int_{\rm{BZ}}d\textbf{k}}
\newcommand{\intbzz}{\int_{\rm{BZ}}d\textbf{k}\int_{\rm{BZ}}d\textbf{k}'}

\usepackage{amsthm}
\theoremstyle{definition}

\begin{document}
\title{Optimally Localized Single-Band Wannier Functions for 2D Chern Insulators}
\author{Thivan M. Gunawardana}
\affiliation{Department of Mathematics, Imperial College London, London SW7 2AZ, United Kingdom}
\author{Ari M. Turner}
\affiliation{Department of Physics, Technion, Haifa 320000, Israel}
\author{Ryan Barnett}
\affiliation{Department of Mathematics, Imperial College London, London SW7 2AZ, United Kingdom}

\begin{abstract}
The construction of optimally localized Wannier functions (and Wannier functions in general) for a Chern insulator has been considered to be impossible owing to the fact that the second moment of such functions is generally infinite. In this article, we propose a solution to this problem in the case of a single band isolated from the rest of the band structure. We accomplish this by drawing an analogy between the minimization of the variance and the minimization of the electrostatic energy of a periodic array of point charges in a smooth neutralizing background. In doing so, we obtain a natural regularization of the diverging variance and this leads to an analytical solution to the minimization problem. We demonstrate our results numerically for a particular model system. Furthermore, we show how the optimally localized Wannier functions provide  a natural way of evaluating the electric polarization for a Chern insulator. 
\end{abstract}		

\maketitle
\section{Introduction}

The quantum mechanics of particles in a periodic potential forms a cornerstone of condensed matter physics. A standard approach in solid-state physics is to work with states that are eigenstates of the underlying Hamiltonian that have crystal momentum as a good quantum number 
\cite{AM}. However, since such Bloch states are extended in real space, they can be ill-suited for understanding local phenomena like covalent bonding in a crystal. Therefore, the construction of Wannier functions \cite{Wannier1}, which are spatially localized states describing a given band or collection of bands, is often desirable. Roughly speaking, these Wannier states bridge the gap between solid-state physics and quantum chemistry, where a local description is preferred.

Since early days, quantum chemists have used so-called Localised Molecular Orbitals (LMO's) to study chemical bonding in molecular systems \cite{qchem1, qchem2}. Wannier functions have provided a powerful means to extend this to studying bonding and other structural properties such as defects in general solid state systems \cite{qchem3, qchem4, qchem5, qchem6}. Furthermore, these states are connected to the Berry phase theory of polarization, providing an alternative and illuminating viewpoint for studying things such as dielectric properties \cite{polarization}. They are also useful as a set of basis states in studying quantum ballistic transport via the Landauer formalism \cite{transport1,transport2}. More recently, Wannier functions have taken a central role in classifying topological band systems \cite{Bradlyn17, Po18}.

To construct Wannier functions for a single band, one integrates the Bloch functions over the Brillouin zone as \cite{Wannier1}
\begin{equation}
 w_{n\textbf{R}}(\textbf{r}) = \frac{1}{V_{\text{BZ}}}\intbz\ e^{-i\textbf{k}\cdot\textbf{R}}\psi_{n\textbf{k}}(\textbf{r}).
    \label{wannier}
\end{equation}
(Details of notation from this section are described more systematically in the following sections.) Such states are typically strongly localized in real space and possess desirable properties: they are orthogonal and the Wannier states for a given Bloch band are simply translations of each other by real space lattice vectors. The above definition has an important ambiguity, namely that the phases of the Bloch states are not specified. One can make the replacement 
$\psi_{n\textbf{k}}(\textbf{r}) \rightarrow e^{i \theta(\textbf{k})} \psi_{n\textbf{k}}(\textbf{r})$, where
$ e^{i \theta(\textbf{k})}$ has $\textbf{k}$-space periodicity, and obtain an equally good definition of the Wannier function. For the multi-band case, there is a larger ambiguity corresponding to a unitary matrix.

In general, changing the gauge in this fashion can have a non-trivial effect on the properties of the Wannier functions. Therefore, it is natural to seek a recipe to the fix the gauge such that we obtain the most desirable Wannier functions. One approach which is well-suited for
conventional (i.e.\ non-topological) systems is to minimize the spatial variance, which directly reflects localization, of the Wannier function with respect to the gauge freedom. This approach, and the resulting Maximally Localized Wannier Functions (MLWFs) \cite{WannierReview2012}, has achieved wide-scale adoption in the electronic structure community and has numerous applications.

A major difficulty emerges \cite{WannierReview2012} when attempting to construct Wannier states for topological insulators such as Chern insulators. Chern insulators are canonical models exhibiting topological bands \cite{Hasan2010, Qi2011}, with the most famous being the Haldane model \cite{Haldane1988}. The difficulty arises due to the presence of a topological obstruction preventing the construction of Bloch states that smoothly depend on ${\bf k}$ while having the correct periodicity. Indeed, if one could find such a smooth dependence, then a straightforward application of Stokes' theorem would reveal that the Chern number will always vanish. We remark that other types of topological obstructions do not necessarily cause issues. For example, it has been shown that globally smooth Bloch states can be constructed in the case of an  $\mathbbm{Z}_2$ obstruction as long as one does not demand that the Wannier functions preserve time-reversal symmetry \cite{soluyanov11}.

To deal with the difficulty in Chern insulators, one approach is to use different gauges to cover two or more patches of the Brillouin zone so that the Bloch states have smooth ${\bf k}$-dependence in each patch. However, such Bloch states will not generally be smooth along lines where the patches meet.
Another approach, which will be taken in this work, is to allow the Bloch states to have vortices in reciprocal space. This approach is sensible because we need to use a single choice of gauge for the Bloch states in (\ref{wannier}), and we would like the Bloch states to be as smooth as possible. Generally, the sum of the winding numbers of all the vortices in the Brillouin zone
will correspond to the Chern number of the band under consideration. For example, for a band with Chern number one, the simplest possibility is to allow for one singly-quantized vortex. Now, the gauge ambiguity allows one a choice not just of local variations of the phase but also of the positions of vortices, which can be moved by singular gauge transformations, which will be discussed in detail later in this article.
Topology dictates the number of defects but not their location.
This phenomenon is similar to the Hairy Ball theorem from algebraic topology \cite{hball}.

The aim of this article is to construct optimally localized Wannier functions for Chern insulating bands. Besides the ambiguity of the vortex positions we just discussed, it is well-known that it is not possible to construct Wannier functions for such systems that decay exponentially at large distances \cite{brouder, Monaco}. This is because we are effectively evaluating the Fourier transform of a function with vortex singularities. The importance of this fact has taken a central role in giving a \textit{defining relation} for topological band systems. In particular, in \cite{Bradlyn17}, a system is defined to be topological if there is no Wannier representation composed of exponentially localized states respecting symmetries of the underlying lattice. In \cite{Po18}, it was argued conversely that states without such an atomic limit are either topological or exhibit so-called fragile topology. The theory of Wannier functions, originally a concept in electronic structure theory, has thus taken a prominent role in the newer field of topological states of matter. Although we cannot construct exponentially localized Wannier functions for a Chern insulator, many of the well-known applications of Wannier functions (e.g. Wannier interpolation and, as will be discussed in this article, the computation of polarization) do not rely on exponential decay. In this article, we construct optimal algebraically localized Wannier functions which are well-suited for these applications.

Wannier functions for a Chern insulating band in 2D will generically decay asymptotically as $1/r^2$ and this is an issue because it leads to a logarithmic divergence of the variance \cite{Monaco}. Additionally, the location of the previously discussed Brillouin zone vortices, which depend on the gauge chosen, will influence quantities which are gauge invariant for the non-topological case such as the so-called Wannier center, which is the average position of a Wannier state. As an example, one can imagine nucleating a vortex and anti-vortex pair infinitesimally close together in the Brillouin zone and then transporting the vortex around the Brillouin zone torus until it 
returns to the anti-vortex and annihilates. Such a process can be achieved by a (singular) gauge transformation alone. This process will continuously shift the Wannier center by one lattice constant. The importance of a systematic way of fixing this singular part of the gauge field is therefore clear.

We solve the problem of minimizing the variance over different vortex positions by identifying a dual picture in which the vortices can be thought of as periodic arrays of point charges in a smooth neutralising background charge distribution. In this setting, minimizing the variance corresponds to minimizing the electrostatic energy per unit cell. This approach provides a natural way to regularize the diverging variance. The source of its divergence is identified with the self-energy of the point charges, which can be neglected (as in ordinary electrostatics) from the total energy. Crucially, this self-energy does not depend on where the point charges are placed. 

To test this theoretical development, we next turn to a concrete example of a Chern insulator lattice model. We construct Wannier functions where the non-singular portions of the gauges are optimised (i.e. are in the Coulomb gauge) but with vortex locations at various places in the Brillouin zone. We demonstrate that the localization of these Wannier functions are in quantitative agreement with the theoretical predictions, with the most localized being the one that minimizes the electrostatic energy in the dual picture. Finally, as a first application of the developed scheme of optimal gauge fixing,
we demonstrate how it can be used to compute the polarization of a Chern insulator, once again illustrating on a particular model.
Even though we have conducted our numerical tests on a particular model, we emphasize that the theoretical results presented in this article are applicable to any Chern insulator.

\section{Background and Preliminary Concepts}
In this section, we give some background to the development of localized Wannier functions and set
notation.
Readers familiar with this may wish to only skim this section.
We begin by considering a periodic quantum system with Hamiltonian $H$, whose eigenstates are given by Bloch's theorem as $\psi_{n\textbf{k}}(\textbf{r}) = e^{i\textbf{k}\cdot\textbf{r}} u_{n\textbf{k}}(\textbf{r})$ where the $u_{n\textbf{k}}(\textbf{r})$ possesses the real-space periodicity of the Hamiltonian. As usual, $n$ labels the band index and $\textbf{k}$ labels reciprocal space wave vectors. We use the normalization convention $\braket{\psi_{n\textbf{k}}}{\psi_{m\textbf{k}'}} = V_{\text{BZ}}\delta_{nm}\delta(\textbf{k} - \textbf{k}')$ for the Bloch states, where $V_{\text{BZ}}$ is the area (or volume in higher dimensions) of the Brillouin zone.

The corresponding Wannier functions \cite{Wannier1} are then defined by (\ref{wannier})
where $\textbf{R}$ denotes a lattice vector. These states are orthonormal and  span the same space as the corresponding Bloch states with which they are constructed. Furthermore, the Wannier functions for a given band are all translations of each other, which can be observed by applying position translation operators to the Wannier functions in the home unit cell.

\subsection{Maximally Localized Wannier Functions}
The primary motivation for Wannier functions comes from the fact they are generally very localized compared to the highly delocalized Bloch states. However, Wannier functions as defined by (\ref{wannier}) are not unique. The Bloch states have an inherent gauge degree of freedom which usually does not have any physical significance. Yet, this choice of gauge has a non-trivial effect on the localization properties of Wannier functions, giving rise to different shapes and decay properties.

To choose between these different Wannier functions, one needs a concrete and systematic way of fixing the gauge of the Bloch states. 
The canonical way of achieving this for non-topological systems is to minimize the total spatial variance
of the set of Wannier functions corresponding to a particular set of Bloch bands as originally described in \cite{Marzari}.
Since the Wannier functions for a given band are translations of the Wannier function in the home unit cell, it is sufficient to minimize the total variance of the Wannier functions in the home unit cell, which is given by the localization functional
\begin{equation}
    F = \sum_{n}\left( \bra{w_{n\textbf{0}}}r^2\ket{w_{n\textbf{0}}} - \bra{w_{n\textbf{0}}}\textbf{r}\ket{w_{n\textbf{0}}}^2 \right),
    \label{locfunreal}
\end{equation}
where $n$ ranges over the set of bands of interest.

It is often more useful to express this variance in momentum space. For the rest of this article, we will focus on a single isolated band and so we drop band indices in what follows. Following work by Blount \cite{Blount}, we can write $F$ in reciprocal space as \cite{Marzari, WannierReview2012}
\begin{equation}
    F = \frac{1}{V_{\text{BZ}}}\intbz\ \braket{\nabla_{\textbf{k}} u_{\textbf{k}}}{\nabla_{\textbf{k}} u_{\textbf{k}}} - \left(\frac{1}{V_{\text{BZ}}}\intbz\ \textbf{A}(\textbf{k})\right)^2,
    \label{locfunrec}
\end{equation}
where $\textbf{A}(\textbf{k}) = i\braket{u_{\textbf{k}}}{\nabla_{\textbf{k}}u_{\textbf{k}}}$ is the Berry connection of the band under consideration (see Appendix \ref{AA} for a detailed derivation). Moreover, we note that $\braket{\nabla_{\textbf{k}} u_{\textbf{k}}}{\nabla_{\textbf{k}} u_{\textbf{k}}} = \braket{\partial_{k_{\alpha}} u_{\textbf{k}}}{\partial_{k_{\alpha}} u_{\textbf{k}}}$, where the Einstein summation convention is used and $\alpha = x,y,z$.

We now wish to minimize the above functional with respect to smooth gauge degrees of freedom of the Bloch states. 
To this end, consider the following family of Bloch states
\begin{equation}
    \ket{u_{\textbf{k}}} = e^{i\theta(\textbf{k})}\ket{\Tilde{u}_{\textbf{k}}},
    \label{gauge}
\end{equation}
where we assume that $\theta(\textbf{k})$ is a smooth and periodic function and 
$\ket{\Tilde{u}_{\textbf{k}}}$ is a fixed reference state.
With respect to this reference state,
the terms in (\ref{locfunrec}) are $\braket{\nabla_{\textbf{k}} u_{\textbf{k}}}{\nabla_{\textbf{k}} u_{\textbf{k}}} = \braket{\nabla_{\textbf{k}} \Tilde{u}_{\textbf{k}}}{\nabla_{\textbf{k}} \Tilde{u}_{\textbf{k}}} - 2(\nabla_{\textbf{k}}\theta(\textbf{k}))\Tilde{\bf A}(\textbf{k}) + (\nabla_{\bf k}\theta(\textbf{k}))^2$ and $\textbf{A}(\textbf{k}) = \Tilde{\textbf{A}}(\textbf{k}) - \nabla_{\textbf{k}}\theta(\textbf{k})$. Then, taking the functional derivative of $F$ with respect to $\theta$ gives 
\begin{equation}
    \frac{\delta F}{\delta \theta} = 2\nabla_{\textbf{k}}\cdot\textbf{A}.
\end{equation}
Thus the maximally localized state will satisfy the Coulomb gauge condition $\nabla_{\textbf{k}}\cdot\textbf{A} = 0$. This quite remarkable and perhaps under-appreciated result was first found by Blount \cite{Blount}. Using this, MLWFs can be constructed numerically via a steepest descent algorithm in such a way that one maintains the unitarity of the gauge transformation. This approach was formulated in \cite{Marzari} and is applicable to the many-band generalization as well and it led to the development of the Wannier90 package for computing MLWFs for real materials \cite{MOSTOFI2008685}.

\subsection{MLWFs for 1D systems}
Let us first consider the 1D case, for completeness and also to demonstrate trouble that arises when extending the treatment to higher dimensions.
Here, it turns out that one can always find exponentially localized Wannier 
functions \cite{kohn}.
Furthermore, it turns out that in 1D, the MLWFs defined previously are simply the eigenstates of the projected position operator $PxP$, where $P$ projects onto the band under consideration. This can be seen quite easily by splitting the variance (\ref{locfunreal}) into
a gauge invariant part $F_I$ and a gauge dependent part $F_G$ as \cite{WannierReview2012}
\begin{align}
    F_I &= \bra{w_{0}} x Q x \ket{w_{0}}
\end{align}
and
\begin{align}
    F_G &= \bra{w_{0}} x P x \ket{w_{0}} - \bra{w_{0}} x \ket{w_{0}}^2\\
    &= \sum_{R\ne 0} | \bra{w_{0}} x \ket{w_{R}}|^2
\end{align}
where the operator $Q = \mathbbm{1} - P$ projects onto the complement of the band under consideration.

Thus, if we  suppose that the $\ket{w_{R}}$s  are eigenstates of $PxP$ with corresponding eigenvalues $\lambda_{R}$, then we find
\begin{equation}
    \bra{w_{0}}x\ket{w_{R}} = \bra{w_{0}}PxP\ket{w_{R}} = \lambda_{0}\delta_{0R}
\end{equation}
and so we see that $F_G = 0$ for these Wannier functions. Since $F_G$ is clearly positive definite and is the only gauge dependent piece, eigenstates of $PxP$ are clearly
maximally localized Wannier states. We note that this argument generalizes straightforwardly to the 1D multi-band case.

\subsection{MLWFs for 2D systems}
Moving to 2D systems, the situation becomes more complicated. We no longer consider eigenstates of the projected position operators because $PxP$ and $PyP$ do not commute in general and are therefore not always simultaneously diagonalizable.
In fact, the commutator of $PxP$ and $PyP$ is directly related to the Chern number
\cite{kitaev06, bianco11, sykes21}.
It was shown in \cite{brouder} that a necessary and sufficient condition for the existence of exponentially localized Wannier functions in 2D is the vanishing of the Chern number of the bands considered. 
Additionally, the variance becomes infinite if the system has non-zero Chern number \cite{Monaco}. As discussed in the introduction, this is due to the presence of irremovable vortices in the Bloch states.
Due to these reasons, one may be led to believe that
MLWFs cannot be constructed in a sensible way for a 2D Chern insulator \cite{thonhauser}.

\section{Theory of Optimally Localized Wannier Functions}
In this section, we present the main result of this article. We provide a solution to the problem of constructing optimally localized Wannier functions for a 2D Chern insulator in the case of a single isolated band. We emphasize that the method described applies even in the case of a complex band structure as long as the band considered is isolated from the rest of the bands.

\subsection{Berry connection and Bloch state vortices}
A Chern insulator is characterized by having a non-zero Chern number, which is defined by \cite{tknn}
\begin{equation}
    C = \frac{1}{2\pi}\intbz\ \Omega(\textbf{k})
\end{equation}
where $\Omega(\textbf{k})$  is the Berry curvature which is gauge invariant. The Chern number is a topological invariant and is always an integer. 
In what follows, we restrict ourselves to the case of
$C = \pm 1$ to avoid unnecessary complexity. However, it is
straightforward to generalize our work to higher Chern
numbers and we will remark on this later.

One usually defines the Berry curvature via $\Omega = \nabla_{\textbf{k}} \times \textbf{A}$ but care must be taken when 
$\textbf{A}$ is not smooth. A standard approach in computing the Chern number is to use different gauges for different patches of the Brillouin zone to ensure $\textbf{A}$ is smooth within each region. Then one simply integrates 
the curl of the smooth connections over each region to evaluate $C$.
On the other hand, in the standard construction of Wannier functions (\ref{wannier}) we require using a 
\textit{single} gauge
to cover the entire Brillouin zone. If one insists on such a gauge, a straightforward application of Stokes' theorem gives 
 \begin{equation}
    \intbz\ \nabla_{\textbf{k}} \times \textbf{A} = 0
    \label{eq:cancel}
\end{equation}
due to the Brillouin zone periodicity of $\textbf{A}$.

There is no contradiction of the above with the possibility of non-zero Chern numbers if we allow
for a vortex in the Bloch states. If there is a vortex at $\textbf{k}_v$, then the Berry connection will also be singular at that point leading to a delta function in its curl. So, we have
\begin{equation}
    \nabla_{\textbf{k}} \times \textbf{A} = \Omega(\textbf{k}) - 2\pi C\delta_P(\textbf{k} - \textbf{k}_v)
\end{equation}
where 
$
\delta_P(\textbf{k}) = \sum_{\textbf{G}}\delta(\textbf{k} + \textbf{G})
$
is the periodic delta function  with summation over reciprocal lattice vectors. Note that the curl of a 2D vector field is shorthand for the scalar quantity $\partial A_y/\partial k_x - \partial A_x/\partial k_y$. Now, (\ref{eq:cancel}) is satisfied because of the cancellation between the delta function and the net Berry curvature. (See Appendix \ref{D} for a related discussion in the context of polarization which will be studied later in this article.)

It is crucial to note that for a Chern insulator, having a vortex is inevitable in this construction. Furthermore,
the position of the vortex is a \textit{gauge dependent} property  because it can be moved by a \textit{singular gauge transformation} i.e. a gauge transformation by a phase with a vortex in it.  If the gauge transformation has a vortex of winding number $-C$ at the initial position and a vortex of winding number $C$ at some other position, then it moves the position of the vortex.
Therefore, the vortex position presents us with an extra degree of freedom to minimize over in addition to the smooth gauge degree of freedom. The latter leads to the Coulomb gauge condition $\nabla_{\mathbf{k}}\cdot\mathbf{A}=0$, which determines the gauge once the position of the vortex is given. Note that for a non-topological system, the Chern number is zero and so there are no vortices to begin with. There, singular gauge transformations should be completely avoided as they would introduce vortices that worsen the decay of the Wannier functions. Thus, one only considers the Coulomb gauge condition in such cases.

\subsection{Electrostatics analogy}
We next proceed to develop the correspondence of the minimization problem with two-dimensional electrostatics.
We begin with the variance functional for a single band in the momentum space representation, as given by (\ref{locfunrec}). Note that this can be broken down into gauge invariant and gauge dependent
parts. In section II.B, we wrote these down in real space form. In momentum space, the gauge dependent part, which is all we need to minimize, takes the form
\begin{equation}
\label{variance2}
\Tilde{F} = \frac{1}{V_{\text{BZ}}}\intbz\ A^2 - \left(\frac{1}{V_{\text{BZ}}}\intbz\ \textbf{A}\right)^2.
\end{equation}
The second term above is gauge invariant under smooth (and periodic) gauge transformations, but we cannot ignore it for the present scenario because it is not invariant under singular gauge transformations.

While the minimization of (\ref{variance2}) over smooth gauge transformations yields the Couloumb gauge condition as usual, this process cannot move the vortex. Therefore, we must solve the Coulomb gauge condition at each vortex position $\textbf{k}_v$ and then use the localization functional to identify the optimal vortex position. In other words, we must solve for $\textbf{A}$ satisfying the equations
\begin{equation}
    \begin{aligned}
    \nabla_{\textbf{k}}\cdot\textbf{A} &= 0 \\
    \nabla_{\textbf{k}}\times\textbf{A} &= \Omega(\textbf{k}) - 2\pi C\delta_P(\textbf{k} - \textbf{k}_v).
    \end{aligned}
    \label{Aeqns}
\end{equation}
These equations uniquely determine ${\bf A}$ up
 to a constant. Furthermore, the integral of
 $\bf A$ over the Brillouin zone \textit{with specified} ${\bf k}_v$ is a gauge invariant quantity which fixes this constant.

\begin{figure*}
    \centering
    \includegraphics[width = 1.5\columnwidth]{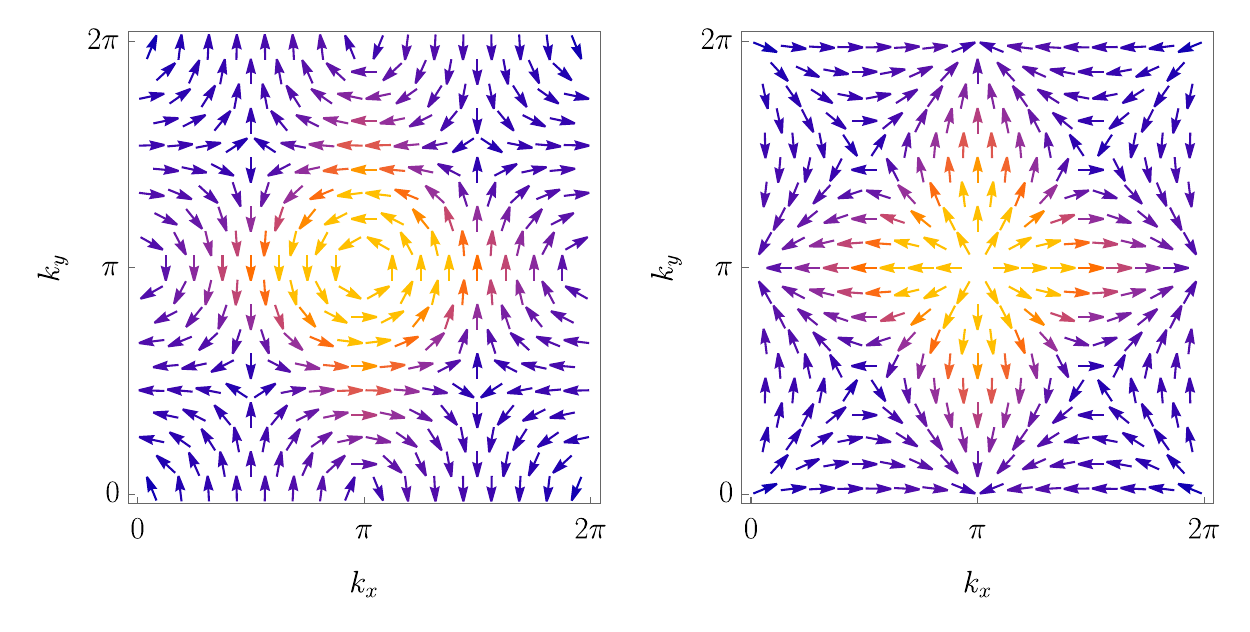}
    \caption{An example of a Berry connection for a set of Bloch states with a vortex (left) and the corresponding zero-average electric field (right). The colors represent the magnitudes of the vectors. Note that there is just one vortex and this is located at $(\pi, \pi)$ (the Berry connection magnitude vanishes at other points where its direction has a defect).}
    \label{vortexelectric}
\end{figure*}

In order to solve (\ref{Aeqns}), we introduce a vector field $\textbf{E}$ such that 
\begin{equation}
\textbf{E} = \hat{\textbf{z}} \times \left({\bf c}- \textbf{A} \right),
\label{Edef}
\end{equation}
where $\textbf{c}=\frac{1}{V_{\text{BZ}}}\intbz\ \textbf{A}$  and $\hat{\textbf{z}}$ is a unit vector perpendicular to the 2D plane on which $\textbf{A}$ lies. Then, in this dual picture, $\textbf{E}$ satisfies the electrostatic equations

\begin{equation}
    \begin{aligned}
      \nabla_{\textbf{k}}\cdot\textbf{E} &= 2\pi\rho(\textbf{k}; \textbf{k}_v) \\
      \nabla_{\textbf{k}}\times\textbf{E} &= 0
    \end{aligned}
    \label{Eeqns}
\end{equation}
with periodic boundary conditions, where $\rho(\textbf{k}; \textbf{k}_v) = \Omega(\textbf{k})/2\pi - C\delta_P(\textbf{k} - \textbf{k}_v)$ is the corresponding charge distribution. In the above, $\textbf{E}$ is interpreted as the electric field of a charge distribution composed of a periodic array of point charges, one per reciprocal space unit cell, together with a smooth background $\Omega$. Note that the net charge per unit cell is zero, which means that this periodic problem is solvable. We emphasize that the Berry connection $\textbf{A}$ should not be confused with an electromagnetic vector potential in the dual picture. Further note that (\ref{Edef}) implies that the average electric field is zero from which it follows that it can be expressed in terms of a periodic potential. Fig. \ref{vortexelectric} provides an example of a Berry connection for a set of Bloch states with a vortex and the corresponding zero-average electric field.

Using (\ref{Edef}) to express the localization functional in terms of ${\bf E}$, we arrive at the simple relation
\begin{equation}
    \Tilde{F} = \frac{1}{V_{\text{BZ}}}\intbz\ E^2.
\end{equation}
The electrostatic analogy therefore goes even further.
In the dual picture, the localization functional becomes the electrostatic energy per unit volume of an infinite periodic charge-neutral system. Minimizing over the singular gauge transformations corresponds to finding the positions of the point charges which minimize this electrostatic energy.

\subsection{Analytical solution}
We now proceed to find
analytical expressions for quantities entering the localization functional.
We split the charge distribution into two parts as $\rho(\textbf{k}; \textbf{k}_v) = \rho_v(\textbf{k}; \textbf{k}_v) + \rho_s(\textbf{k})$ where
\begin{equation}
    \rho_v(\textbf{k}; \textbf{k}_v) = \rho_0 - C\delta_P(\textbf{k} - \textbf{k}_v)
\end{equation}
corresponds to a periodic lattice of point charges in a \textit{uniform} neutralizing background given by $\rho_0 = C/V_{\text{BZ}}$ and
\begin{equation}
    \rho_s(\textbf{k}) = \frac{\Omega(\textbf{k})}{2\pi} - \rho_0
\end{equation}
corresponds to a smooth periodic charge distribution in the same uniform neutralizing background.

We will solve (\ref{Eeqns}) for these two charge distributions separately and thus obtain a solution to the entire problem by linearity. For the charge distribution $\rho_v(\textbf{k}; \textbf{k}_v)$, one typically uses the Ewald summation technique to find the electric field \cite{ewald, Tosi1964, Bar08}. In the present problem, we are in 2D so we can instead leverage results from complex analysis. For each wave vector $\textbf{k}$, we introduce a corresponding complex representation given by $z = k_x + ik_y$. Furthermore, we represent the electric field $\textbf{E}_v(\textbf{k}; \textbf{k}_v)$ for this problem in complex form as $E_v(z; z_v) = E_x - iE_y$, where $z_v$ is the complex representation of the vortex position $\textbf{k}_v$. Then, following ideas from work by Tkachenko on superfluid vortex lattices \cite{tkachenko}, we find that the solution to the electric field is given by
\begin{equation}
    E_v(z; z_v) = -C[\zeta(z - z_v) + \alpha (z-z_v)] + \rho_0\pi (z^*-z^*_v),
    \label{Ev}
\end{equation}
where $\alpha$ is a constant chosen to make $E_v$ periodic (see Appendix \ref{B} for more details) and $\zeta(z)$ is the Weierstrass zeta function given by 
\begin{equation}
     \zeta(z) = \frac{1}{z} + \sum_{G\neq 0}\left(\frac{1}{z - G} + \frac{1}{G} + \frac{z}{G^2}\right).
\end{equation}
Here, $G = G_x + iG_y$ is the complex representation of reciprocal lattice vector $\textbf{G}$. 

For this electric field, the corresponding electrostatic potential is given by
\begin{equation}
    \begin{aligned}
        \phi_v(z; z_v) = \frac{1}{2}C\log\abs{\sigma(z- z_v)}^2 + C\bigg(\frac{1}{4}&\alpha (z-z_v)^2 + \text{c.c.}\bigg) \\
        &- \frac{1}{2}\rho_0\pi\abs{z-z_v}^2
    \end{aligned}
    \label{potential}
\end{equation}
where $\sigma(z)$ is the Weierstrass sigma function which is related to $\zeta(z)$ via 
\begin{equation}
    \zeta(z) = \frac{d}{dz}\log\sigma(z).
\end{equation}
We note that this electrostatic potential is both real and periodic (see Appendix \ref{B}).

The solution is much simpler for the smooth charge distribution $\rho_s(\textbf{k})$.
We can write the Fourier series representation
\begin{equation}
    \rho_s(\textbf{k}) = \frac{1}{2\pi}\sum_{\textbf{R}\neq 0} \Omega_{\textbf{R}} e^{-i\textbf{k}\cdot\textbf{R}},
\end{equation}
where the $\Omega_{\textbf{R}}$s are the Fourier components of the Berry curvature.
Note that there is no ${\bf R}=0$ term in the above equation due to charge neutrality of $\rho_s(\mathbf{r})$.
By taking a Fourier transform of Poisson's equation, we therefore obtain the solutions
\begin{equation}
    \textbf{E}_s(\textbf{k}) = \sum_{\textbf{R}\neq 0}\frac{i\textbf{R}}{R^2}\Omega_{\textbf{R}}e^{-i\textbf{k}\cdot\textbf{R}}
\end{equation}
and
\begin{equation}
    \phi_s(\textbf{k}) = \sum_{\textbf{R}\neq 0} \frac{\Omega_{\textbf{R}}}{R^2}e^{-i\textbf{k}\cdot\textbf{R}}
\end{equation}
for the electric field and electrostatic potential respectively.

The solution to the full problem (\ref{Eeqns}) is obtained by summing together the two solutions $\textbf{E}_v$ and $\textbf{E}_s$ we found above. Note that the total potential $\phi = \phi_v + \phi_s$ is periodic in reciprocal space so the electric field obtained indeed has zero-average.

\subsection{Minimizing $\Tilde{F}$}
We now proceed to describe the minimization of the localization functional.
Using the solution found in Section III.C, we can solve the problem of minimizing the localization functional $\Tilde{F}$. In the electrostatics picture, we have
\begin{align}
    \Tilde{F} &= \frac{1}{V_{\text{BZ}}}\intbz\ \norm{\textbf{E}_v + \textbf{E}_s}^2 \nonumber \\
    &= \frac{1}{V_{\text{BZ}}}\left(\intbz\ E_v^2 -2\intbz\ \textbf{E}_v\cdot\nabla_{\textbf{k}}\phi_s + \intbz\ E_s^2\right) \nonumber \\
    &= \text{const} - \frac{4\pi C}{V_{\text{BZ}}}\phi_s(\textbf{k}_v),
    \label{minresult}
\end{align}
where the constant term does not depend on the vortex position. The non-constant term follows from integrating by parts the term involving the interaction between the point particle and the smooth charge distribution.

Note that this immediately gives us a sensible way to regularize the blowing up of the variance mentioned previously. In the electrostatics picture, this infinity simply corresponds to the self energy of the point charges, which is contained in the term involving $E_v^2$. This does not depend on where the charges are placed and so it can be neglected in the usual fashion. In the variance picture, this term corresponds to a logarithmic divergence which is independent of the vortex position. Therefore, we can regularize the localization functional by introducing a long distance cutoff and thus effectively minimize the finite short-range contribution to the localization functional over the vortex position. This will be discussed further in the next 
section.
Moreover, (\ref{minresult}) tells us that to find optimal Wannier functions for $C =  +1 (-1)$ in our setting, the vortex needs to be placed at the maximum (minimum) of the electrostatic potential $\phi_s$ for the smooth charge distribution in a uniform neutralizing background. 

As stated in the beginning of this section, the generalization to higher Chern numbers is relatively straightforward.
For a Chern number $|C|=1$, there is only one vortex per unit cell. The vortex-vortex interaction contribution to the electrostatic energy will not depend on the vortex position because changing the vortex position corresponds to uniformly translating the vortex lattice. Thus we have lumped the vortex-vortex interaction into the constant term in (\ref{minresult}). This, however, does not remain true for $|C| \ge 2$ where there will be two or more vortices in the Brillouin zone. Here, one must explicitly consider the vortex-vortex interaction terms involving (\ref{potential}) in the minimization which makes the problem richer.  For this case, there will generally be two competing terms in 
(\ref{minresult}):
one describing vortex-vortex interaction and the other describing the vortices interacting with the smooth potential.

\section{Numerical Results}
\begin{figure}
    \centering
    \includegraphics[width = 0.9\columnwidth]{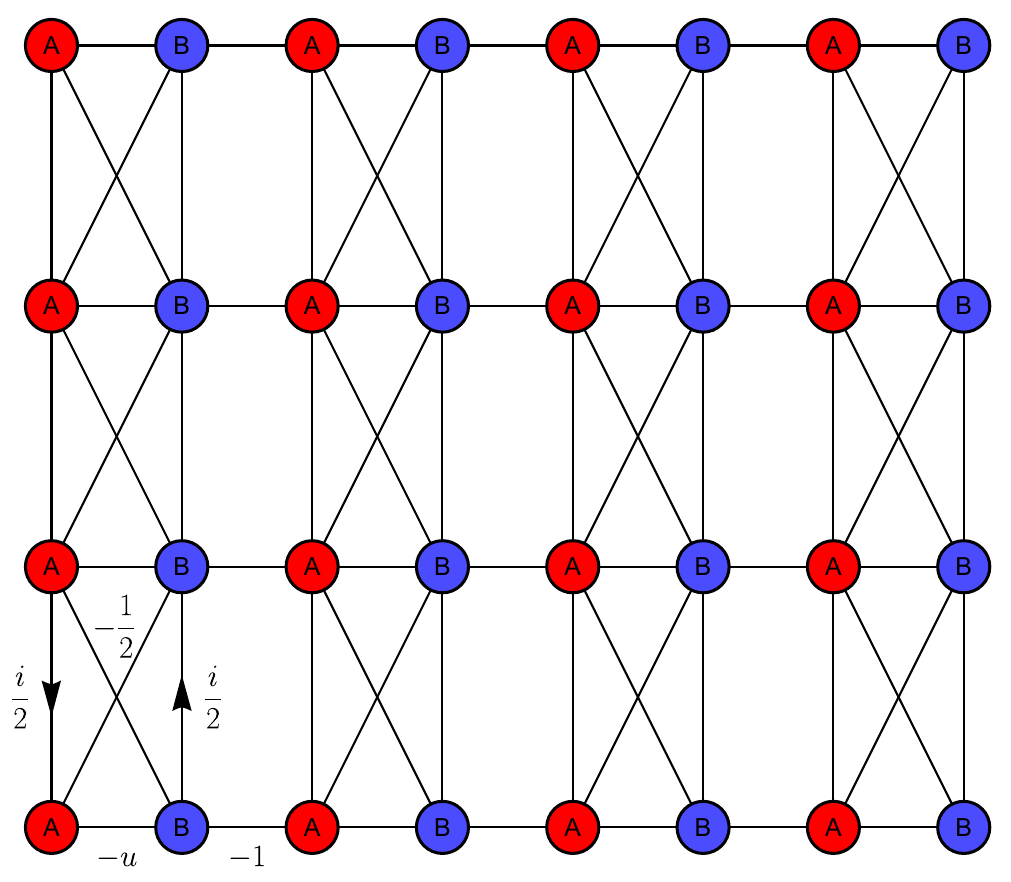}
    \caption{Diagram of the real space lattice connected to the Hamiltonian given by (\ref{rhamiltonian}). Sites on sublattices A and B are colored red and blue respectively. The arrows indicate the hopping terms.}
    \label{latticemodel}
\end{figure}

In this section, we present numerical results to test the developments of the previous section. 
We use a minimal Chern insulator model that is related to the
Qi-Wu-Zhang model \cite{qwz} via a spin rotation. The lattice Hamiltonian considered is
\begin{equation}
\begin{aligned}
    H = &-\sum_{i,j} a^{\dagger}_{i+1,j}b_{i,j} + \text{h.c.} \\
    &- \frac{1}{2}\sum_{i,j}(a^{\dagger}_{i,j}b_{i,j+1} + a^{\dagger}_{i,j}b_{i,j-1}) + \text{h.c.} \\
    &- \frac{i}{2}\sum_{i,j}(a^{\dagger}_{i,j+1}a_{i,j} + b^{\dagger}_{i,j}b_{i,j+1}) + \text{h.c.} \\
    & -u\sum_{i,j} a^{\dagger}_{i,j}b_{i,j} + \text{h.c.}
\end{aligned}
\label{rhamiltonian}
\end{equation}
This is a Hamiltonian on a lattice with sublattices $A$ and $B$ where the creation operators $a^{\dagger}_{i,j}$ and $b^{\dagger}_{i,j}$ create a particle at the $A$ and $B$ sites of the $(i,j)$ unit cell respectively. We set the lattice constants to one so that the distance between the $A$ and $B$ sites in a given unit cell is 1/2. The Hamiltonian involves nearest-neighbor hoppings in the $x$ and $y$ directions and also cross hoppings between $A$ and $B$ sites of different unit cells, as shown in Fig. \ref{latticemodel}. One of the hoppings is imaginary in order to break time-reversal symmetry (similar to the Haldane model's complex next-nearest-neighbor hoppings).

Since this Hamiltonian is periodic in real space, one can Fourier transform to obtain the following Bloch Hamiltonian
\begin{equation}
\begin{aligned}
    H(\textbf{k}) = &-(1 + u + \cos(k_y))\cos\left(\frac{k_x}{2}\right)\sigma_x \\
    &-(1 - u - \cos(k_y))\sin\left(\frac{k_x}{2}\right)\sigma_y \\
    &- \sin(k_y)\sigma_z,
\end{aligned}
\label{rhamiltonian}
\end{equation}
where $\sigma_{x,y,z}$ are the Pauli matrices.
Note that we have chosen to Fourier transform the Hamiltonian in such a way that it is not periodic in reciprocal space. This has been done to incorporate the different positions of the $A$ and $B$ sublattice sites within the unit cell into the real space position operator. This choice of position operator is much more intuitive for our discussion since the Wannier representation is an inherently real space quantity.
The Hamiltonian (\ref{rhamiltonian}) has lowest-band Chern number $C = 1$ for $0 < u < 2$, $C = -1$ for $-2 < u < 0$ and $C = 0$ for $\abs{u} > 2$. For the numerical work in this section, we set $u = 1$ so that the Chern number is always +1.
Throughout we will consider Wannier functions of the lower band.

\begin{figure}
\subfloat[]{
\includegraphics[width = 0.5\columnwidth]{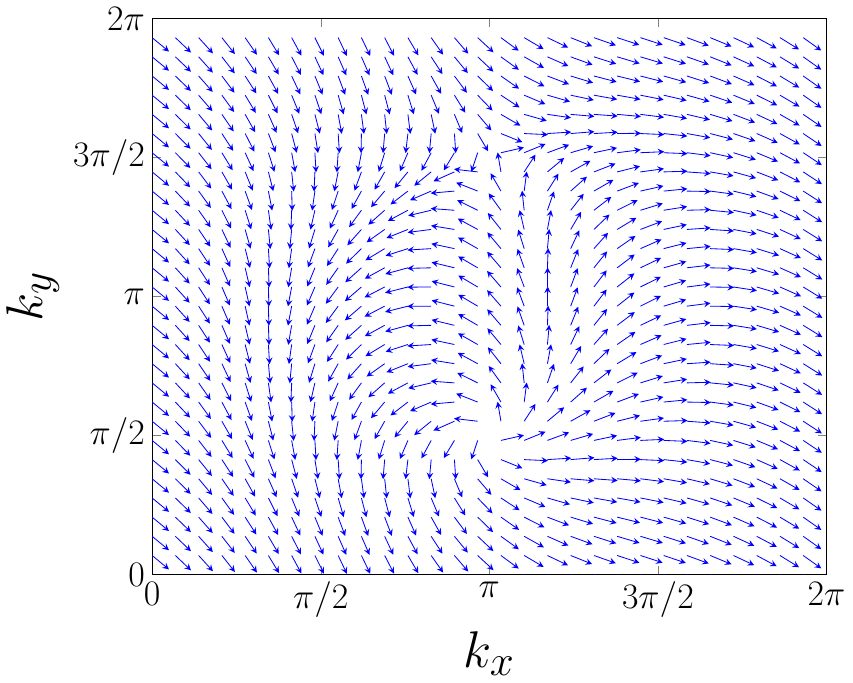}
}
\subfloat[]{
\includegraphics[width = 0.5\columnwidth]{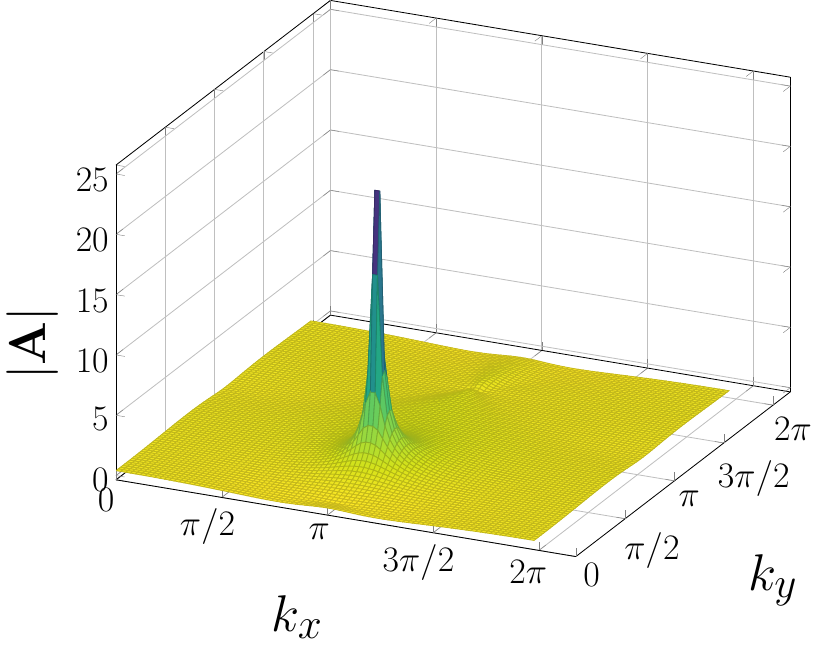}
}\\
\subfloat[]{
\includegraphics[width = 0.5\columnwidth]{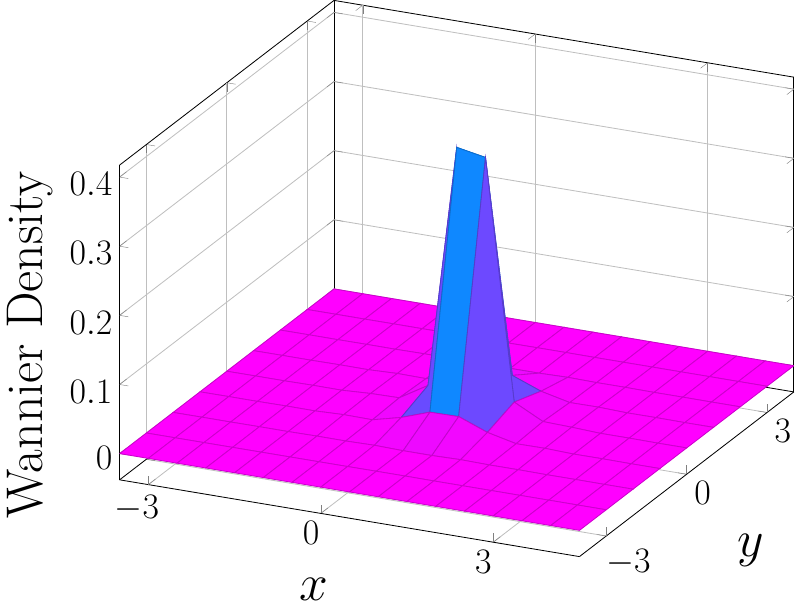}
}
\subfloat[]{
\includegraphics[width = 0.5\columnwidth]{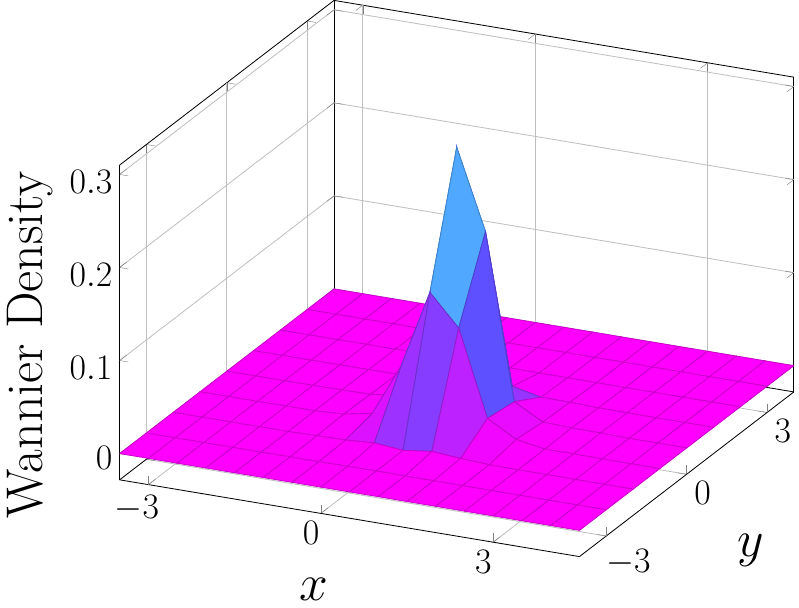}
}\\
\subfloat[]{
\includegraphics[width = 0.5\columnwidth]{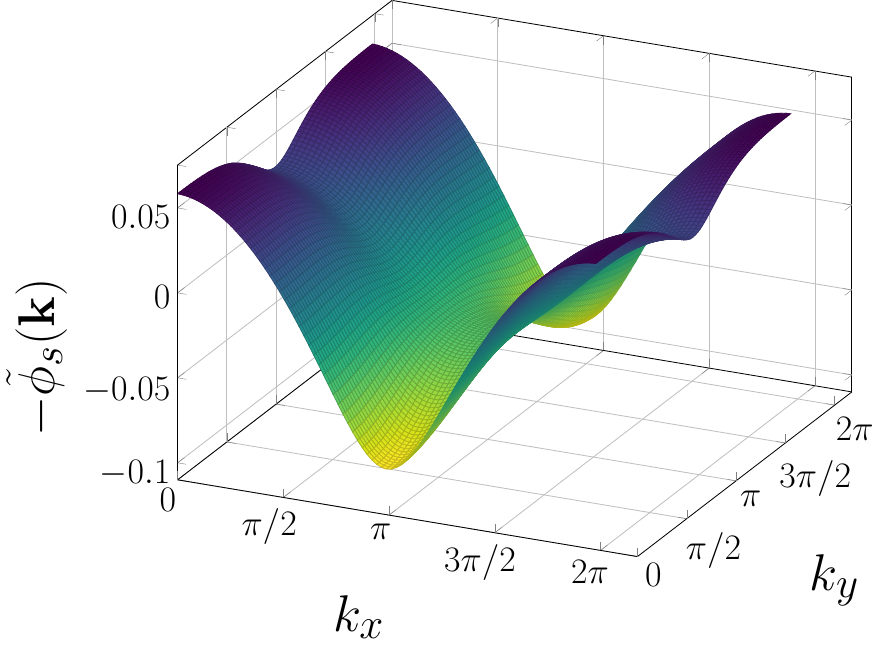}
}
\subfloat[]{
\includegraphics[width = 0.5\columnwidth]{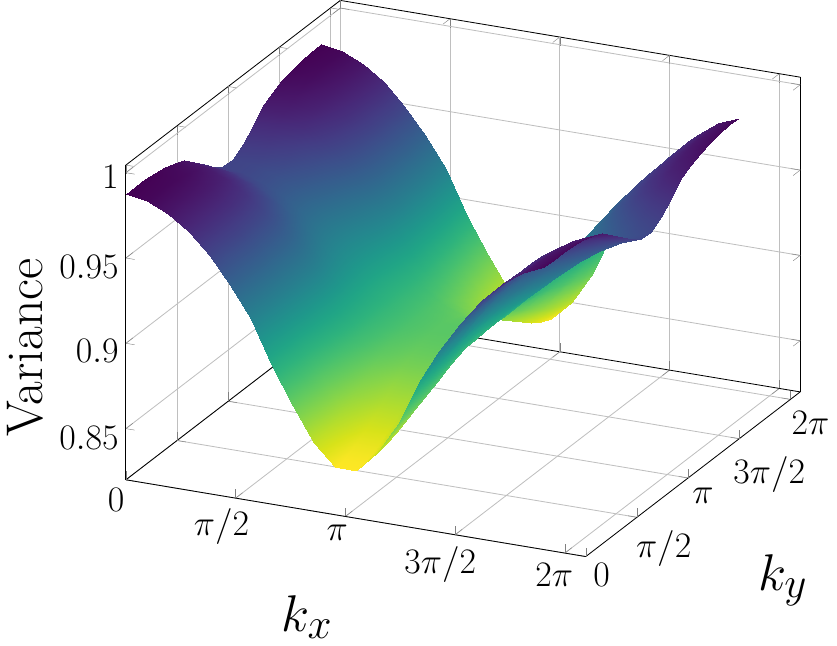}
}
\caption{\textbf{Numerical Results for the Lattice Model.} $N = 87$ is used everywhere except for the vector plot. (a) Vector plot of the singular gauge transformation for $N = 29$ which moves the vortex from $(\pi, 3\pi/2)$ to $(\pi, \pi/2)$. The arrows are the vectors formed by the real and imaginary parts of this gauge transformation. (b) Magnitude of the Berry connection after applying the singular gauge transformation from (a). (c) Zoomed-in plot of the optimal Wannier function in the home unit cell. This has a modified variance (see text) of 0.8461. (d) Zoomed-in plot of the Wannier function obtained by optimising over smooth gauges but placing the vortex at the sub-optimal position $(\pi/4, \pi/4)$. This has a modified variance (see text) of 0.9758. (e) Plot of minus the modified smooth potential $\Tilde{\phi}_s(\textbf{k}) = (4\pi/V_{\text{BZ}})\phi_s(\textbf{k})$. (f) Plot of the modified variance (see text) of the optimal Wannier function obtained by placing the vortex in different points across the Brillouin zone. The number of different vortex positions used is 441.}
\label{quadplot}
\end{figure}

In order to compute optimal Wannier functions for this model, we begin by discretizing the Brillouin zone into $N$ equally spaced points in each direction. Moreover, we need to make an initial choice of gauge for the Bloch states. Let us begin by fixing the first entry of the
eigenvector of $H({\bf k})$ to always be real.
 It can then be seen  that this fixes the vortex in the position $(k_x, k_y) = (\pi, 3\pi/2)$. 
 
 We now require a general singular gauge transformation capable of shifting this vortex to an arbitrary location in the Brillouin zone. Such a gauge transformation is not unique but we can fix it by also demanding that it leaves the value of $\nabla_{\textbf{k}}\cdot\textbf{A}$ unchanged. This is a useful choice because it allows us to carry out minimizations over smooth and singular gauge transformations in any order. It follows directly from the solution to the electrostatic problem in the previous section (see also Appendix \ref{C}) that the singular gauge transformation shifting a vortex from position $a$ to $b$ (where both are written in complex representation), is given by $e^{i\theta_s(z; a,b)}$, where
\begin{equation}
\begin{aligned}
    \theta_s(z; a,b) = -C\Im\bigg(\log&\left(\frac{\sigma(z-a)}{\sigma(z-b)}\right) + \alpha(b-a)z \\
    &\qquad \ \ + \frac{\pi}{V_{\text{BZ}}}(b^* - a^*)z\bigg).
\end{aligned}
\label{singuage}
\end{equation}
Here, $z = k_x + ik_y$ is once again the complex representation of the reciprocal space vector $\textbf{k}$ and $\alpha$ is the same constant as in (\ref{Ev}). This gauge transformation satisfies $\nabla^2\theta_s = 0$ by construction (thus preserving $\nabla_{\textbf{k}}\cdot\textbf{A}$) and 
$e^{i\theta_s(z; a,b)}$ is Brillouin zone periodic.
 Fig. \ref{quadplot} (a) shows a vector plot of the singular gauge transformation for $N = 29$ and $a = \pi + 3\pi i/2$ and $b = \pi + \pi i/2$. The arrows represent the direction of the phase $e^{i\theta_s}$ on the unit circle. We observe that this gauge transformation creates an anti-vortex at $a$ and a vortex at $b$, thereby completely shifting the vortex from $a$ to $b$ without leaving any trace behind.

To confirm the above, Fig. \ref{quadplot} (b) shows the magnitude of the Berry connection in the Brillouin zone after the vortex has been shifted using the gauge transformation in (a). Here, $N = 87$ is used. Note that a vortex in the ground state wavefunction at some point $\textbf{k}_v$ in the Brillouin zone leads to the behavior $|\textbf{A}(\textbf{k})| \sim 1/|\textbf{k} - \textbf{k}_v|$ close to the vortex and we can therefore conclude from (b) that our gauge transformation has shifted the vortex correctly.

Next, we compute the optimal Wannier functions for our lattice model using the analytical result from the previous section. Namely, noting that we have $C = +1$, we first move the vortex to the maximum of the smooth potential $\phi_s(\textbf{k})$ and then minimize the variance functional with respect to smooth gauge transformations. The latter is done via a steepest descent algorithm. In our single band case, this amounts to solving the differential equation
\begin{equation}
    \frac{dU(\textbf{k}, \tau)}{d\tau} = -i(\nabla_{\textbf{k}}\cdot\textbf{A}(\textbf{k}, \tau))U(\textbf{k}, \tau)
\end{equation}
where $U(\textbf{k}, \tau) = e^{i\theta(\textbf{k}, \tau)}$ is the smooth unitary gauge transformation acting on the Bloch states and $\tau$ is a suitably scaled relaxation time variable. Note that we always time-evolve $U$ in such a way that unitarity is preserved at each step. It is crucial to note that the steepest descent process cannot alter the vortex position by definition. Fig. \ref{quadplot} (c) shows the density of the resulting optimal Wannier function in real space. In contrast, Fig. \ref{quadplot} (d) shows the Wannier function for a gauge satisfying $\nabla\cdot\textbf{A} = 0$ but with the vortex at $(\pi/4, \pi/4)$, which is sub-optimal. It is clear that this Wannier function overlaps non-trivially with a neighboring unit cell and so the importance of choosing the right vortex position is evident.

We now move on to consider the variance of the optimally localized Wannier functions we have found.
For an unbiased check of the analytical arguments,
we will endeavor to compute the variance working entirely within real space. 
This involves finding a real-space method of regularizing the divergence of the variance.
As we have seen, in $\bf k$-space the divergence is regularized simply by dropping the vortex self-interaction term. In real space, we expect that the variance should be regularized by long-distance modifications. 
The variance in real space can be written in the form
\begin{equation}
    F = \int d\textbf{r} \int d\textbf{r}' |w(\textbf{r})|^2|w(\textbf{r}')|^2V(\textbf{r} - \textbf{r}'),
    \label{genlocfun}
\end{equation}
where $w(\textbf{r})$ is the Wannier function and $V(\textbf{r}) = r^2/2$. 
(The idea of expressing localization functionals in this form is not new - two such examples are the Edmiston-Ruedenberg criterion \cite{qchem1} and the von Niessen criterion \cite{vniessen}.)
The regularization procedure then 
is to alter the ``potential" $V(\textbf{r})$  at large distances to achieve a convergent result.  
We work with the potential given by
\begin{equation}
    \Tilde{V}(\textbf{r}) = R\sqrt{R^2 + r^2} - R^2,
\end{equation}
where $R$ is a cutoff parameter. We observe that $\Tilde{V}(\textbf{r}) \sim r^2/2$ for $r \ll R$ while $\Tilde{V}(\textbf{r}) \sim Rr$ for $r \gg R$. Thus, this choice of potential has the desired properties, given the $1/r^2$ decay of the Wannier functions. 
For our calculations, we require a value of $R$
that is large enough to capture the finite part of the variance. This can be done by choosing $R$ such that $R^2$ is large compared to the ratio of the band curvature and the band gap, at $\textbf{k} = 0$. Thus, we now have a sensible choice of cut-off for the localization functional which can be applied to arbitrary system sizes (which are larger than the cut-off). We will refer to this version of the localization functional as the \textit{modified variance}.

With this, we can verify that the optimal Wannier functions are indeed found by placing the vortex in the maximum of the potential and that the dependence of the localization functional on the vortex position is of the form given by (\ref{minresult}). We carry out the steepest descent procedure while placing the vortex in different positions across the Brillouin zone. For each vortex position, we evaluate the modified variance of the resulting Wannier function using a cutoff radius of $R = 30$. The result is shown in Fig. \ref{quadplot}(f). We compare this with Fig. \ref{quadplot}(e), which is a plot of the negative of the modified smooth potential $\Tilde{\phi}_s$ (see (\ref{minresult})), which is defined to be
\begin{equation}
    \Tilde{\phi}_s = \frac{4\pi}{V_{\text{BZ}}}\phi_s.
\end{equation}
We see that these two plots differ only by an additive constant as predicted by (\ref{minresult}), thus validating our analytical result.

\section{Application: Polarization}

Having developed and tested optimally localized Wannier functions for Chern insulators, we now move on to describe an application. In particular, we will show how they can be used to compute the polarization of a Chern insulator in an unambiguous and physically relevant way.

For topologically trivial systems, Wannier functions provide an elegant means of computing the electric polarization, following the modern theory of polarization \cite{polarization, vanderbilt_book}. The fundamental idea is that an absolute polarization cannot be defined uniquely for an extended system, but the change in polarization of a system subject to some deformation is well-defined and experimentally measurable. It can also be shown that this change in polarization is given by the change in the position of the Wannier center in real space, providing a very intuitive  picture \cite{WannierReview2012}.

Polarization for a Chern insulator, on the other hand, has been a topic under some amount of debate, with recent works studying this extensively 
(see, for example, \cite{song21, barkeshli23, vaidya23}). Indeed, the polarization is conventionally determined by integrating the Berry connection over the Brillouin zone. However, as we have discussed, this quantity is not invariant with respect to singular gauge transformations. Therefore, a consistent mechanism of fixing the singular gauge is crucial. We will show that the construction we have for optimally localized Wannier functions provides a very natural way of fixing the gauge and thereby obtain an expression for the polarization in terms of the Wannier center shift plus a correction. Our results are fully consistent with those of \cite{Coh} 
where a ${\bf k}$-space expression for the polarization of a Chern insulator is developed.

Polarization is encapsulated by the
fundamental relation  \cite{polarization}
\begin{equation}
    \Delta\textbf{P}_n = \int_0^1 d\lambda\ \textbf{J}_n(\lambda),
    \label{poldef}
\end{equation}
which describes the change in polarization as the integral of the current during the deformation process. Here, $\Delta\textbf{P}_n$ is the change in polarization of the $n^{\text{th}}$ band and $\lambda$ parametrizes the adiabatic deformation of the Hamiltonian ($\lambda = 0$ is usually the centrosymmetric case). Furthermore,
\begin{equation}
    \textbf{J}_n(\lambda) = -\frac{ie}{(2\pi)^2}\intbz\ \braket{\partial_{\lambda}u_{n\textbf{k}}}{\nabla_{\textbf{k}}u_{n\textbf{k}}} + \text{c.c.}
\end{equation}
is the cell-averaged adiabatic current where $-e$ is the charge quantum, and we note that the Bloch states are now functions of $\lambda$. This adiabatic current is a measurable and therefore a gauge invariant quantity, thus providing the correct starting point for considering polarization in a Chern insulator.

We now recall the Berry connection $\textbf{A}_{n\textbf{k}} = i\braket{u_{n\textbf{k}}}{\nabla_{\textbf{k}}u_{n\textbf{k}}}$  and introduce 
$A_{n\lambda} = i\braket{u_{n\textbf{k}}}{\partial_{\lambda}u_{n\textbf{k}}}$. In terms of these, we may write the adiabatic current as
\begin{equation}
\begin{aligned}
    \textbf{J}_n(\lambda) = -\frac{e}{(2\pi)^2}&\intbz\ \bigg(\partial_{\lambda}\textbf{A}_{n\textbf{k}} - \nabla_{\textbf{k}}A_{n\lambda} \\
    &+ i\bra{u_{n\textbf{k}}}(\nabla_{\textbf{k}}\partial_{\lambda} - \partial_{\lambda}\nabla_{\textbf{k}})\ket{u_{n\textbf{k}}}\bigg).
\end{aligned}
\end{equation}
The important term here is the final one. Mixed partial derivatives commute when acting on a smooth function but for a Chern insulator, they would not commute at vortices and so we cannot ignore this term. Therefore, for a band with $C = \pm 1$, we find that (see Appendix 
\ref{D})
\begin{equation}
       \Delta\textbf{P}_n = -\frac{e }{V_c} \Delta\overline{\textbf{r}}_n + \frac{eC}{2\pi}\hat{\textbf{z}}\times\Delta\textbf{k}_{vn},
    \label{chernpol}
\end{equation}
where $\Delta\overline{\textbf{r}}_n$ is the change in the Wannier center, $\Delta\textbf{k}_{vn}$ is the change in the vortex position of the Bloch states due to the adiabatic deformation, and $V_c$ is the area of the unit cell. This relation
(\ref{chernpol}) is another main result of the article.
For conventional (non-topological) systems, the correct expression for the polarization is given by the first term of  (\ref{chernpol}) only. However, this term is not invariant with respect to singular gauge transformations and the additional term in  (\ref{chernpol})  serves to cancel off this gauge dependence, making the full expression of the polarization gauge invariant as it must be.

\begin{figure}
\includegraphics[width = \columnwidth]{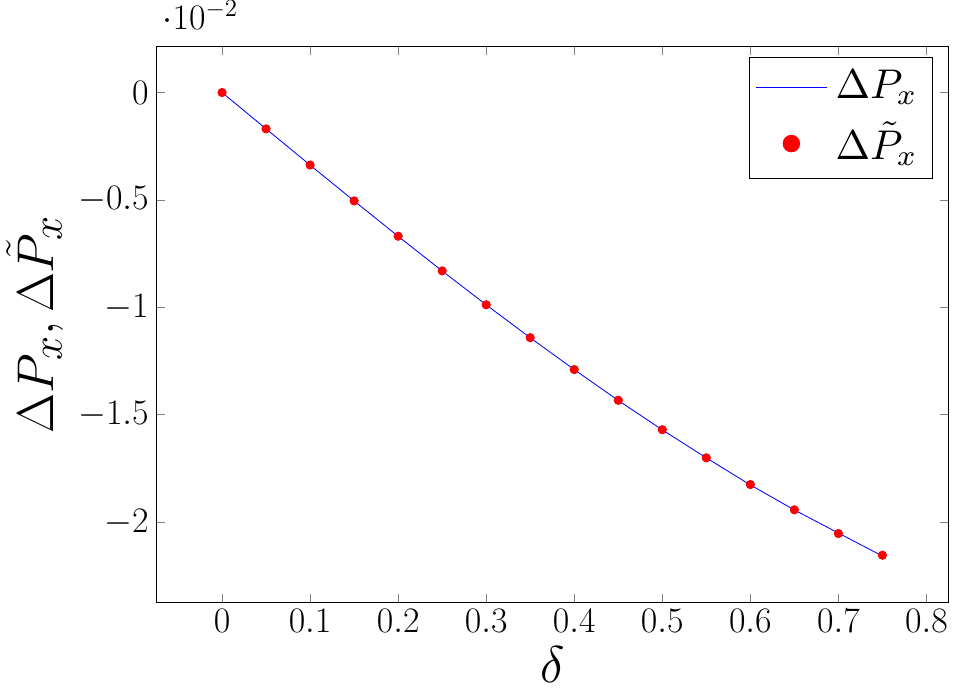}
\caption{Plots of change in polarization (blue) for the lattice model with respect to the centrosymmetric case and change in dipole moment (red dots) across an infinite strip of the lattice model. Both quantities are given in units of $-e/a$ where $a$ is the lattice constant. A reciprocal space grid with $N = 80$ was used to compute the polarization. To compute the dipole moment, Richardson extrapolation was applied with two strip widths $M = 20$ and $M = 30$ and two reciprocal space discretisations with 500 and 1000 $k$-points respectively.}
\label{polarization}
\end{figure}

We now wish to use the developed formalism to compute the polarization of our model system. In order to induce a change in polarization, we add a staggered onsite potential which breaks centrosymmetry and so we consider the modified Hamiltonian
\begin{equation}
    \Tilde{H} = H + \delta\sum_{i,j}(a^{\dagger}_{i,j}a_{i,j} - b^{\dagger}_{i,j}b_{i,j}).
\end{equation}
where $\delta$ is the onsite potential. We set $u = 0.5$ so that the hopping amplitudes alternate in strength in the $x$-direction.
Starting at the centrosymmetric case $\delta = 0$, we compute the change in polarization as the staggering is adiabatically increased up to $\delta = 0.75$.  We note that we cannot cross $\delta = \sqrt{3}/2 \approx 0.866$ since the gap would close, breaking adiabaticity. Throughout we fix the average particle number to be one particle per every two lattice sites.
By considering the symmetry of Fig. \ref{latticemodel}, we see that turning on $\delta$ would only induce a polarization in the $x$-direction.  We use optimally localized Wannier functions in each case to compute the polarization via (\ref{chernpol}), where we use an $80 \times 80$ grid of points in reciprocal space. The results, which can be computed fairly efficiently, are shown by the blue curve in Fig. \ref{polarization}.

To check the validity of the expression for the polarization,
we now turn to numerically computing the dipole moment of our model for large finite system sizes, mostly  following the method described in \cite{Coh}.
We take the system to have open
boundary conditions in the $x$-direction (which is the direction of the induced polarization) and periodic boundary conditions in the $y$-direction.
For a system with $M$ unit cells along the $x$-direction, the dipole moment per unit length is given by
\begin{equation}
    \Tilde{P}_x = -\frac{e}{MV_c}\int_0^{2\pi}dk\ \sum_{\substack{n\in\text{occupied}\\ \text{states}}}\ \bra{\psi_{nk}}x\ket{\psi_{nk}}
    \label{dipolemoment}
\end{equation}
where $V_c$ is the area (volume in higher dimensions) of the real space unit cell, $n$ is the band index and $k$ is the wave number. Note that $k$ is now a scalar because we only have periodicity in the $y$-direction, and the number of bands involved scales with $M$.

There is then a question of how to fill the edge states when computing the dipole moment as $\delta$ is varied. Here we appeal to a sensible physical scenario. 
We start with the zero-temperature ground state when $\delta=0$. 
We take the rate of variation to be slow with respect to the bulk energy gap, but fast with respect to edge-to-edge tunneling which exponentially decays as a function of system length. This means that, during the course of the evolution, the population of edge states may differ from those of instantaneous ground states.
Specifically, the final state will acquire a different chemical potential on each side through
the quasi-adiabatic evolution. This is how occupied states in
(\ref{dipolemoment})  are determined.

The change in dipole moment with respect to $\delta = 0$, calculated as prescribed above, is shown by red dots in Fig. \ref{polarization}. We compute the dipole moment for two widths $M = 20$ and $M = 30$, and two $k$-space discretizations with 500 and 1000 $k$-points. We then apply Richardson extrapolation in both $M$ and number of $k$-points to eliminate the leading order errors in both quantities. This gives us an excellent approximation for the dipole moment in the thermodynamic limit $M \rightarrow \infty$, so that it may be compared with the polarization. Indeed, we observe very good agreement as seen in the plot, where the largest relative error is only $0.17\%$ and occurs for $\delta = 0.75$, which is in the vicinity of where the gap closes.

\section{Summary and Outlook}
In this article, we have proposed a means of computing optimally localized Wannier functions for a single band in Chern insulators. We showed that the problem of minimizing the variance functional for such systems has a dual interpretation involving minimizing the electrostatic energy of a periodic array of point charges in a smooth neutralising background. In this dual picture, the electrostatic energy naturally involved an infinite self-energy term. We observed that if this term were neglected in the usual fashion, then we obtained a finite expression which depended non-trivially on the positions of the vortices. Minimizing this finite part corresponds to minimizing the short range spread of the Wannier function around its center so that the overlap with neighboring atomic sites is minimized. This makes our optimal Wannier functions useful in practical calculations. We used Chern numbers of $\pm 1$, where there is always only one vortex, for clarity and showed that the vortex position minimizing the electrostatic energy was an extremum of the electrostatic potential corresponding to the (neutralized) smooth background. We then remarked that everything works similarly for higher Chern numbers in the case of a single band - the electrostatic energy simply involves extra terms due to pairwise vortex interactions (since there can be more than one vortex) and it must be minimized over all over vortex arrangements. 

As an immediate application of our work, we presented a Wannier function approach to computing polarization in a Chern insulator. In doing so, we showed that the fact that the Wannier center is no longer gauge invariant in the usual sense is not problematic given that we start with a physical gauge invariant quantity (namely the adiabatic current). Moreover, we verified our result by comparing it with the dipole moment per unit length which is also a physical quantity.

Our work leaves many avenues open for future research. Foremost among them is the generalization to multi-band settings. There, one must minimize the localization functional over all gauge transformations within that subspace of bands. This includes singular non-Abelian gauge transformations which dictate the distribution of the vortices across the multiple bands. Furthermore, multi-band problems often involve band crossings which give rise to further discontinuities. In the non-topological case, such discontinuities can be removed using smooth non-abelian gauge transformations \cite{WannierReview2012}. We speculate that it may be possible to do the same for band crossings when the total Chern number is non-zero. If this is possible, then we can pick a gauge in which one of the new bands (no longer eigenstates of the Hamiltonian) has all the vortices while the rest of the new bands are smooth. In that case, all but one of the Wannier functions would be exponentially localized and so it may be possible that such a scenario leads to the optimal choice. An additional generalization is the case of a smaller subspace of states disentangled from a larger set of entangled bands (for example by restricting to states within a specific energy window). This is discussed for conventional systems in \cite{WannierReview2012}.

Another issue to consider is that of Wannier functions displaying spatial symmetries of the underlying Hamiltonian. This is a topic that has been studied in depth in many different settings since the early days \cite{cloizeaux63, kohn73, strinati78, ku02}. In particular, there have been examples such as the s-like Wannier function of Cu \cite{souza01}, where the optimal Wannier function does not exhibit the right symmetries of the system. As a result, methods have been introduced to obtain maximally localized symmetric Wannier functions for non-topological systems by extending the methods of Marzari and Vanderbilt \cite{sakuma13}. It remains to be studied whether the optimally localized Wannier functions we propose preserve key symmetries or whether our methods can similarly be extended to obtain symmetric Wannier functions by trading off some amount of localization. Indeed, to study this fully, one must first extend our work to multiple bands. Moreover, the fact the Wannier center in our case can be shifted arbitrarily via singular gauge transformations may be crucial to this discussion.

Other potential applications of our work remain to be investigated. For example, our optimal Wannier functions may be used to construct minimal tight-binding models for Chern insulating bands in a similar fashion to techniques applied in topologically trivial settings. Finally, how the Wannier construction described in this article applies to so-called fragile topological phases \cite{Po18} warrants further consideration.

\begin{acknowledgements}
We are grateful to Arash Mostofi, Darren Crowdy, Peru d'Ornellas and Frank Schindler for discussions and helpful comments.
TMG acknowledges support from an Imperial Department of Mathematics Roth PhD studentship. AMT acknowledges support from the Israeli Science
Foundation (ISF) Grant No. 1939/18. RB acknowledges support from an Imperial Mathematics Research Impulse Grant and the Aspen Center for Physics under NSF Grant No. PHY-2210452 where part of this work was completed.
\end{acknowledgements}

\appendix
\section{Derivation of the localization functional in reciprocal space}
\label{AA}

In this appendix, we present a derivation of the localization functional in reciprocal space given by (\ref{locfunrec}) in the main text. We begin with the matrix elements of the position operator in the Bloch basis, which are given by \cite{Blount}
\begin{equation}
    \bra{\psi_{n\textbf{k}}}\textbf{r}\ket{\psi_{m\textbf{k}'}} = iV_{\text{BZ}}(\delta_{nm}\nabla_{\textbf{k}} + \braket{u_{n\textbf{k}}}{\nabla_{\textbf{k}}u_{m\textbf{k}}})\delta(\textbf{k} - \textbf{k}').
    \label{blountformula}
\end{equation}
Using the above, we can immediately find
\begin{align}
    \bra{w_{n\textbf{0}}}\textbf{r}\ket{w_{n\textbf{0}}} &= \frac{1}{V_{\text{BZ}}^2}\intbzz\ \bra{\psi_{n\textbf{k}}}\textbf{r}\ket{\psi_{n\textbf{k}'}} \nonumber \\
    &= \frac{1}{V_{\text{BZ}}}\intbz\ \textbf{A}_{n\textbf{k}}.
\end{align}

Moreover, we note that
\begin{equation}
    \bra{\psi_{n\textbf{k}}}r^2\ket{\psi_{n\textbf{k}'}} = \bra{\psi_{n\textbf{k}}}\textbf{r}\cdot P\textbf{r}\ket{\psi_{n\textbf{k}'}} + \bra{\psi_{n\textbf{k}}}\textbf{r}\cdot Q\textbf{r}\ket{\psi_{n\textbf{k}'}},
    \label{decomp}
\end{equation}
where $P = \frac{1}{V_{\text{BZ}}}\intbz\ \ket{\psi_{n\textbf{k}}}\bra{\psi_{n\textbf{k}}}$ projects onto the band under consideration and $Q = \mathbbm{1} - P$ projects onto the complement of this band.

We can compute each term on the RHS of (\ref{decomp}) separately using (\ref{blountformula}). This gives
\begin{align}
    \bra{\psi_{n\textbf{k}}}\textbf{r}\cdot P\textbf{r}\ket{\psi_{n\textbf{k}'}} &= \frac{1}{V_{\text{BZ}}}\intbz''\  \bra{\psi_{n\textbf{k}}}\textbf{r}\ket{\psi_{n\textbf{k}''}}\cdot \bra{\psi_{n\textbf{k}''}}\textbf{r}\ket{\psi_{n\textbf{k}'}} \nonumber \\
    &= V_{\text{BZ}}\intbz''\ (i\nabla_{\textbf{k}} + \textbf{A}_{n\textbf{k}})\delta(\textbf{k} - \textbf{k}'') \nonumber \\
    &\qquad\qquad \cdot (i\nabla_{\textbf{k}''} + \textbf{A}_{n\textbf{k}''})\delta(\textbf{k}'' - \textbf{k}') \nonumber \\
    &= V_{\text{BZ}} (i \nabla_{\textbf{k}} + {\bf A}_{n {\bf k}})^2 \delta({\bf k}-{\bf k}')
    \label{rPr}
\end{align}
and
\begin{align}
    \bra{\psi_{n\textbf{k}}}\textbf{r}\cdot Q\textbf{r}\ket{\psi_{n\textbf{k}'}} &= \frac{1}{V_{\text{BZ}}}\sum_{l\neq n}\intbz''\  \bra{\psi_{n\textbf{k}}}\textbf{r}\ket{\psi_{l\textbf{k}''}}\cdot \bra{\psi_{l\textbf{k}''}}\textbf{r}\ket{\psi_{n\textbf{k}'}} \nonumber \\
    &= V_{\text{BZ}}\sum_{l\neq n}\braket{\nabla_{\textbf{k}}u_{n\textbf{k}}}{u_{l\textbf{k}}}\cdot\braket{u_{l\textbf{k}}}{\nabla_{\textbf{k}}u_{n\textbf{k}}}\delta(\textbf{k} - \textbf{k}'), 
    \label{rQr}
\end{align}
where we used the fact that $\braket{u_{n\textbf{k}}}{\nabla_{\textbf{k}}u_{l\textbf{k}}} = -\braket{\nabla_{\textbf{k}}u_{n\textbf{k}}}{u_{l\textbf{k}}}$.
Summing together (\ref{rPr}) and (\ref{rQr}) and noting that $\sum_l \ket{u_{l\textbf{k}}}\bra{u_{l\textbf{k}}} = \mathbbm{1}$, we obtain
\begin{align}
  \bra{\psi_{n\textbf{k}}}r^2\ket{\psi_{n\textbf{k}'}} = V_{\text{BZ}}\bigg(-\nabla_{\textbf{k}}^2 &+ 2 i\textbf{A}_{n\textbf{k}}\cdot \nabla_{\bf k}+ i \nabla_{\bf k} \cdot
  \textbf{A}_{n\textbf{k}} \nonumber \\
  &+ \braket{\nabla_{\textbf{k}}u_{n\textbf{k}}}{\nabla_{\textbf{k}}u_{n\textbf{k}}}\bigg)\delta(\textbf{k} - \textbf{k}'),
\end{align}
from which it follows that
\begin{equation}
    \bra{w_{n\textbf{0}}}r^2\ket{w_{n\textbf{0}}} = \frac{1}{V_{\text{BZ}}}\intbz\ \braket{\nabla_{\textbf{k}}u_{n\textbf{k}}}{\nabla_{\textbf{k}}u_{n\textbf{k}}}.
\end{equation}

\section{Summary of special functions used}
\label{A}
In this appendix to be self-contained, we provide a short summary of the Weierstrass functions, 
which are special functions central to our solution of the electrostatics problem. Tailoring to the problem of interest, we cast everything in reciprocal space.
The classic text Ref. \cite{watson} is recommended for additional details.

Elliptic functions are doubly periodic functions in the complex plane and their only singularities are a finite number of poles per unit cell. In order to connect this directly to our problem, let us cast the 2D reciprocal space in complex form, as discussed in the main text. That is, we define $z = k_x + ik_y$ to be the complex number representing the reciprocal wave vector $\textbf{k} = (k_x, k_y)$. Then, we can identify Brillouin zone periodicity with double periodicity in the complex representation.

With this setup, the Weierstrass $\wp$-function in complex reciprocal space is defined as
\begin{equation}
    \wp(z) = \frac{1}{z^2} + \sum_{G \neq 0}\left(\frac{1}{(z-G)^2} - \frac{1}{G^2}\right),
\end{equation}
where $G = G_x + iG_y$ is the complex representation of reciprocal lattice vector \textbf{G}. It can be seen that this function is even  and it has poles at reciprocal lattice vectors. Moreover, let the complex primitive reciprocal lattice vectors be denoted $2\omega_1$ and $2\omega_2$. Then, it can be shown that $\wp(z)$ is doubly periodic with half-periods $\omega_1$ and $\omega_2$. (Here, we choose $\omega_1$ and $\omega_2$ such that $\omega_2/\omega_1$ has positive imaginary part.) In other words, we have
\begin{equation}
    \wp(z + 2\omega_1) = \wp(z + 2\omega_2) = \wp(z).
    \label{p_periodicity}
\end{equation}
Therefore, $\wp(z)$ is an elliptic function and it has the required Brillouin zone periodicity.

This function, however, is not immediately helpful in solving the electrostatics problem from the main text
  (\ref{Eeqns})
as we require simple, not double poles. Instead, we use the Weierstrass zeta function which is defined by
\begin{equation}
    \frac{d\zeta}{dz} = -\wp(z)
    \label{zetadef}
\end{equation}
together with the condition
\begin{equation}
    \lim_{z\rightarrow 0}\left(\zeta(z) - \frac{1}{z}\right) = 0.
\end{equation}
By term-by-term integration, it can be shown that the $\zeta$-function has the series representation
\begin{equation}
    \zeta(z) = \frac{1}{z} + \sum_{G \neq 0}\left(\frac{1}{z-G} + \frac{1}{G} + \frac{z}{G^2}\right)
    \label{zetaseries}
\end{equation}
which exhibits the desired simple poles.

In contrast to the $\wp$-function, the $\zeta$-function is not elliptic because it is only quasi-periodic in the Brillouin zone. 
To see this, one can first  integrate
(\ref{zetadef}) to find
\begin{equation}
    \zeta(z+2\omega_i) = \zeta(z) + 2\eta_i,
    \label{zetaperiodicity}
\end{equation}
where the last term is a constant of integration. Next, note that
$\zeta(z)$ is an odd function as can be observed from its series expansion.
Substituting $z=-\omega_i$ into \ref{zetaperiodicity} now allows us to determine the integration constants $\eta_i = \zeta(\omega_i)$ which are generally non-zero.

We now state an important relation between the $\omega_i$'s and the $\eta_i$'s, namely
\begin{equation}
    \eta_1\omega_2 - \eta_2\omega_1 = \frac{i\pi}{2}
    \label{etaomega}
\end{equation}
This can be derived by taking a contour integral of $\zeta(z)$ around the boundary of a unit cell. Using the residue theorem together with the fact the integrals
along opposite edges are related via periodicity,
we can obtain (\ref{etaomega}). (This result relies on our assumption that $\omega_2/\omega_1$ has positive imaginary part since this determines the handedness of the integration contour.)

The final special function we  require, for the purpose of constructing the electromagnetic potential, is the Weierstrass sigma function, which is best defined via its relationship to the $\zeta$-function as
\begin{equation}
    \frac{d}{dz}(\log\sigma(z)) = \zeta(z),
    \label{sigmadiffeqn}
\end{equation}
together with the extra condition
\begin{equation}
    \lim_{z\rightarrow 0}\frac{\sigma(z)}{z} = 1.
\end{equation}
Using similar techniques as for the $\zeta$-function, we can show that the $\sigma$-function is also not doubly periodic but instead satisfies the quasi-periodic relations
\begin{equation}
    \begin{aligned}
        \sigma(z + 2\omega_1) = -e^{2\eta_1(z+\omega_1)}\sigma(z), \\
        \sigma(z + 2\omega_2) = -e^{2\eta_2(z+\omega_2)}\sigma(z).
    \end{aligned}
    \label{sigmaperiodicity}
\end{equation}

\section{Detailed Solution of Electrostatics Problem}
\label{B}
In this appendix, we provide more details for the solution of the electrostatic equations for a lattice of point charges in a uniform neutralizing background. We recall that the electrostatic equations to be solved are given by
\begin{equation}
    \begin{aligned}
        \nabla_{\textbf{k}}\cdot\textbf{E} &= 2\pi(\rho_0 - C\delta_P(\textbf{k}-\textbf{k}_v))\\
        \nabla_{\textbf{k}}\times\textbf{E} &= 0,
        \label{electrostatics}
    \end{aligned}
\end{equation}
where $\textbf{k}_v$ is the position of the vortex in the Brillouin zone.

The electric field for a single point charge in an infinite 2D plane is given according to Coulomb's law as $\textbf{E}(\textbf{k}; \textbf{k}_v) = (\textbf{k} - \textbf{k}_v)/|\textbf{k} - \textbf{k}_v|^2$. In the complex representation, this electric field is given by $E(z) = E_x - iE_y = 1/(z - z_v)$ where $z = k_x + ik_y$ and $z_v$ is the point charge position as usual. Moving to the periodic case, we can try to extend the point charge solution in the most obvious way as
\begin{equation}
    \Tilde{E}(z; z_v) = \sum_G \frac{1}{z-z_v-G}
    \label{pointcharge}
\end{equation}
where $G$ represents reciprocal lattice vectors.
This sum, however, does not converge which is problematic.

The solution is to use the Weierstrass $\zeta$-function introduced in the previous section. Informally, we can understand the $\zeta$-function as a regularization of (\ref{pointcharge}) - that is, it introduces additional terms which make the new sum convergent outside of the poles.
Despite fixing the divergence, and having poles in the correct locations, the $\zeta$-function is not doubly periodic. 
This can be dealt with by adding terms linear in $k_x$ and $k_y$.
In particular, a direct calculation using relations from the previous Appendix shows that the expression
\begin{equation}
    E(z;z_v) = -C[\zeta(z-z_v) + \alpha (z-z_v)] + \pi\rho_0 (z^*-z_v^*),
    \label{electricfield}
\end{equation}
where
\begin{equation}
    \alpha = \frac{\eta_1\omega_2^* - \eta_2\omega_1^*}{\omega_1^*\omega_2 - \omega_2^*\omega_1} = \frac{2i(\eta_2\omega_1^* - \eta_1\omega_2^*)}{V_{\text{BZ}}},
    \label{alpha}
\end{equation}
has full Brillouin-zone periodicity. A similar expression was obtained in early work on vortex lattices in superfluid helium \cite{tkachenko}. The quantity $\alpha$ depends solely on the lattice geometry and for a square lattice it can be found that $\alpha = 0$. This can be observed by noting that $\omega_2 = i\omega_1$ and using (\ref{zetaseries}) to show that $\eta_2 = \zeta(\omega_2) = -i\zeta(\omega_1) = -i\eta_1$.

We now show that the electric field found indeed satisfies the required equations (\ref{electrostatics}).
Using the identity
$d/dz^*((z-z_v)^{-1}) = \pi\delta(\textbf{k} - \textbf{k}_v)$, and noting the $\zeta$ function has simple poles,
one finds
\begin{align}
    (\nabla_{\textbf{k}}\cdot\textbf{E}) - i(\nabla_{\textbf{k}}\times\textbf{E}) = 2\frac{dE}{dz^*}
    &= 2 \pi( \rho_0 -C \delta_P({\bf k}- {\bf k}_v)).
    \label{Echeck2}
\end{align}

Noting that {\bf E} is real,
the equations (\ref{electrostatics}) are 
satisfied.

Finally, as stated in the main text, the electrostatic potential for this problem is given by
\begin{equation}
    \begin{aligned}
        \phi_v(z; z_v) = \frac{1}{2}C\log\abs{\sigma(z- z_v)}^2 + C\bigg(\frac{1}{4}&\alpha (z-z_v)^2 + \text{c.c.}\bigg) \\
        &- \frac{1}{2}\rho_0\pi\abs{z-z_v}^2
    \end{aligned}
    \label{vortexpotential}
\end{equation}
This satisfies $\textbf{E}(\textbf{k}) = -\nabla_{\textbf{k}}\phi_v$ since $-2(d\phi_v/dz) = E(z)$. This is also periodic as claimed, which can be seen by using the quasi-periodic conditions (\ref{sigmaperiodicity}) for the $\sigma$-function.

\section{Derivation of Singular Gauge Transformation}
\label{C}
In this appendix, we derive Equation (\ref{singuage}) from the main text, which is the formula for the singular gauge transformation which shifts the vortex position from one point in the Brillouin zone to another while preserving the Coulomb gauge condition.

We begin with the fact that under a gauge transformation of the form $\ket{u_{n\textbf{k}}} \rightarrow e^{i\theta(\textbf{k})}\ket{u_{n\textbf{k}}}$, the Berry connection transforms as $\textbf{A} \rightarrow \textbf{A} - \nabla_{\textbf{k}}\theta$. If we move the vortex from position \textbf{a} to \textbf{b}, then the Berry connection initially and finally must satisfy
\begin{equation}
    \begin{aligned}
        \nabla_{\textbf{k}}\times\textbf{A} &= \Omega(\textbf{k}) - 2\pi C\delta_P(\textbf{k} - \textbf{a}), \\
        \nabla_{\textbf{k}}\times\Tilde{\textbf{A}} &= \Omega(\textbf{k}) - 2\pi C\delta_P(\textbf{k} - \textbf{b})
    \end{aligned}
    \label{curlAab}
\end{equation}
respectively.
If we demand the additional condition that this gauge transformation keeps $\nabla_{\textbf{k}}\cdot\textbf{A}$ unchanged (so that we can shift the vortex after minimizing over continuous gauge transformations in one position without having to carry out steepest descent again), then it follows from the above that $\textbf{v} = \nabla_{\textbf{k}}\theta$ must satisfy
\begin{equation}
    \begin{aligned}
        \nabla_{\textbf{k}}\times \textbf{v} &= 2\pi C(\delta_P(\textbf{k} - \textbf{b}) - \delta_P(\textbf{k} - \textbf{a})), \\
        \nabla_{\textbf{k}}\cdot \textbf{v} &= 0.
    \end{aligned}
    \label{periodpotfloweqns}
\end{equation}
If we now define $\textbf{v} = \hat{\textbf{z}} \times \textbf{E}_{\theta} + \text{const.}$, then (\ref{periodpotfloweqns}) also turns into an electrostatics problem which can be solved by using the Weierstrass $\zeta$-function. Based on our solution to the electrostatics problem in the main text and Appendix \ref{B}, we can write down the solution for the complex electric field $E_{\theta}$ as
\begin{equation}
    E_{\theta}(z; a,b) = C(\zeta(z-b) - \zeta(z-a) + S)
\end{equation}
where $S$ is a constant.
Based on this, the solution for $\textbf{v}$ in complex form may be written as
\begin{equation}
    v(z; a,b) = iC(\zeta(z-a) - \zeta(z-b) + \Tilde{S})
    \label{vsol}
\end{equation}
where $v=v_x-iv_y$ and $\Tilde{S}$ is a different constant.

Our aim is to obtain an expression for $\theta$ from (\ref{vsol}). To achieve this, we note that (\ref{periodpotfloweqns}) are fluid equations for a periodic incompressible potential flow around a vortex and an anti-vortex. The complex velocity is a meromorphic function of $z$.
So using methods familiar from fluid dynamics in 2D, we proceed by introducing a complex potential $f = \theta + i\psi$ where $\theta$ is the velocity potential, corresponding precisely to our gauge function, and $\psi$ is the stream function. This complex potential is defined to satisfy
\begin{equation}
    v = \frac{df}{dz}.
\end{equation}
It then follows from (\ref{vsol}) and (\ref{sigmadiffeqn}) that
\begin{equation}
    f(z; a,b) = iC\left(\log\left(\frac{\sigma(z-a)}{\sigma(z-b)}\right) + \Tilde{S}z\right).
    \label{fsol}
\end{equation}
Note that we have thrown out the constant of integration in (\ref{fsol}) because this would simply correspond to a global phase, which we can ignore.

Now, our singular gauge $\theta(z; a,b)$ is given by the real part of (\ref{fsol}) and we require that $e^{i\theta}$ is periodic in the Brillouin zone. We now use this condition to obtain the constant $\Tilde{S}$. It follows from (\ref{sigmaperiodicity}) that
\begin{equation}
    \begin{aligned}
            \theta(z + 2\omega_i; a,b) = \theta(z;a,b) + iC\bigg(&\eta_i(b-a) - \eta_i^*(b^* - a^*)\\
            &+ \Tilde{S}\omega_i - \Tilde{S}^*\omega_i^*\bigg) + 2n_i\pi
    \end{aligned}
\end{equation}
for $i=1,2$, where the $n_i$'s are integers. Note that the $2n_i\pi$ term arises due to the nature of the complex logarithm. Thus, for periodicity of $e^{i\theta}$, it is sufficient that
\begin{equation}
    \eta_i(b-a) - \eta_i^*(b^* - a^*) + \Tilde{S}\omega_i - \Tilde{S}^*\omega_i^* = 0
    \label{Seqns}
\end{equation}
for $i = 1,2$.
Using (\ref{etaomega}) and the fact that $\omega_1^*\omega_2 - \omega_2^*\omega_1 = iV_{\text{BZ}}/2$, we solve equations (\ref{Seqns}) to obtain
\begin{equation}
    \Tilde{S} = \frac{\pi}{V_{\text{BZ}}}(b^*-a^*) + \alpha(b-a)
\end{equation}
where $\alpha$ is the same constant as in (\ref{alpha}).
Hence we finally obtain
\begin{equation}
\begin{aligned}
    \theta(z; a,b) = -C\Im\bigg(\log&\left(\frac{\sigma(z-a)}{\sigma(z-b)}\right) + \alpha(b-a)z \\
    &\qquad \ \ + \frac{\pi}{V_{\text{BZ}}}(b^* - a^*)z\bigg),
\end{aligned}
\label{singuageapp}
\end{equation}
as required. We emphasize that while $\theta(z;a,b)$ may change by integer multiples of $2\pi$ when translated by reciprocal lattice vectors, $e^{i\theta(z;a,b)}$ is Brillouin-zone periodic.

\section{Polarization}
\label{D}
In this appendix, we provide a detailed derivation of equation (\ref{chernpol}) for the change in polarization in a Chern insulator with $C = \pm 1$. We recall once again that the change in polarization is defined by \cite{polarization}
\begin{equation}
    \Delta\textbf{P}_n = \int_0^1 d\lambda\ \textbf{J}_n(\lambda),
    \label{poldefapp}
\end{equation}
where
\begin{equation}
    \textbf{J}_n(\lambda) = -\frac{ie}{(2\pi)^2}\intbz\ \braket{\partial_{\lambda}u_{n\textbf{k}}}{\nabla_{\textbf{k}}u_{n\textbf{k}}} + \text{c.c.}
\end{equation}
is the cell-averaged adiabatic current.  Note that all the quantities in this expression are functions of $\lambda$.

This can be written in terms of the Berry connection $\textbf{A}_{n\textbf{k}} = i\braket{u_{n\textbf{k}}}{\nabla_{\textbf{k}}u_{n\textbf{k}}}$ and the new quantity $A_{n\lambda} = i\braket{u_{n\textbf{k}}}{\partial_{\lambda}u_{n\textbf{k}}}$ as
\begin{equation}
\begin{aligned}
    \textbf{J}_n(\lambda) = -\frac{e}{(2\pi)^2}&\intbz\ \bigg(\partial_{\lambda}\textbf{A}_{n\textbf{k}} - \nabla_{\textbf{k}} A_{n\lambda} \\
    &+ i\bra{u_{n\textbf{k}}}(\nabla_{\textbf{k}}\partial_{\lambda} - \partial_{\lambda}\nabla_{\textbf{k}})\ket{u_{n\textbf{k}}}\bigg).
\end{aligned}
\label{current}
\end{equation}
It is easy to see that the first term in the current gives the change in Wannier center when integrated over $\lambda$, while integrating the second term gives zero due to periodicity of the Berry connection. Let us now focus on the last term in the current. This clearly vanishes except at the vortex itself. Therefore, let us consider a small disk $D_{\epsilon}$ of radius $\epsilon$ around the vortex. In such a region, we may write 
\begin{equation}
    \ket{u_{n\textbf{k}}} = e^{i\theta(\textbf{k} - \textbf{k}_{vn}(\lambda))}\ket{\Tilde{u}_{n\textbf{k}}},
    \label{nearvortex}
\end{equation}
where $\ket{\Tilde{u}_{n\textbf{k}}}$ is smooth and $\textbf{k}_{vn}(\lambda)$ is the vortex position. 
Thus, in our small region, we have extracted the vortex out of the Bloch state and into a phase factor. (It is not possible to factor out a phase from $\ket{\Tilde{u}_{n\textbf{k}}}$ over the whole Brillouin zone without creating a new vortex somewhere else, as vortices cannot be eliminated from a band with a Chern number, but we will use this expression only within $D_\epsilon$.)

In $D_{\epsilon}$, it is true that
\begin{equation}
    \nabla_{\textbf{k}}\times\textbf{A}_{n\textbf{k}} = \nabla_{\textbf{k}}\times\Tilde{\textbf{A}}_{n\textbf{k}} - \nabla_{\textbf{k}}\times\nabla_{\textbf{k}}\theta.
\end{equation}

So, we have
\begin{align}
            0 &= \intbz\ \nabla_{\textbf{k}}\times\textbf{A}_{n\textbf{k}} \nonumber \\
            &= \int_{\text{BZ}\backslash D_{\epsilon}}d\textbf{k}\ \nabla_{\textbf{k}}\times\textbf{A}_{n\textbf{k}} +\int_{D_{\epsilon}} d\textbf{k}\ \nabla_{\textbf{k}}\times\Tilde{\textbf{A}}_{n\textbf{k}} \nonumber \\
            &\qquad\qquad\qquad\qquad\qquad\ \ \ \ - \int_{D_{\epsilon}} d\textbf{k}\ \nabla_{\textbf{k}}\times\nabla_{\textbf{k}}\theta \nonumber \\
            &= \intbz\ \Omega(\textbf{k}) - \int_{D_{\epsilon}}d\textbf{k}\ \nabla_{\textbf{k}}\times\nabla_{\textbf{k}}\theta
\end{align}
where the first equality follows from Stokes' theorem and the periodicity of $\textbf{A}_{n\textbf{k}}$, and the third equality follows since $\nabla_{\textbf{k}} \times \textbf{A}_{n\textbf{k}}$ is equal to the gauge invariant Berry curvature $\Omega_{n}(\textbf{k})$ in regions of $\textbf{k}$-space free of vortices. Similar arguments were used implicitly in section III.A of the main text. Hence, we obtain that
\begin{equation}
    \int_{D_{\epsilon}}d\textbf{k}\ \nabla_{\textbf{k}}\times\nabla_{\textbf{k}}\theta = C
\end{equation}
$\forall\epsilon > 0$, and thus taking the limit as $\epsilon\rightarrow 0$, we deduce that
\begin{equation}
    \nabla_{\textbf{k}}\times\nabla_{\textbf{k}}\theta = 2\pi C\delta_P(\textbf{k} - \textbf{k}_{vn}(\lambda)).
    \label{curlgrad}
\end{equation}

Returning back to (\ref{current}), we can use the chain rule to obtain
\begin{align}
    \partial_{\lambda}\theta(\textbf{k}-\textbf{k}_{vn}(\lambda)) &= \nabla_{\textbf{k}}\theta(\textbf{k}-\textbf{k}_{vn}(\lambda))\cdot(-\partial_{\lambda}\textbf{k}_{vn}), \\
    \partial_{\lambda}\nabla_{\textbf{k}}\theta(\textbf{k}-\textbf{k}_{vn}(\lambda)) &= (-\partial_{\lambda}\textbf{k}_{vn})\cdot\nabla_{\textbf{k}}\bigg[\nabla_{\textbf{k}}\theta(\textbf{k}-\textbf{k}_{vn}(\lambda))\bigg]
\end{align}
and therefore,
\begin{align}
&\bra{u_{n\mathbf{k}}}(\partial_x\partial_{\lambda} - \partial_{\lambda}\partial_x)\ket{u_{n\mathbf{k}}} \nonumber\\
   =\ &i(\partial_x\partial_{\lambda} - \partial_{\lambda}\partial_x)\theta(\textbf{k}-\textbf{k}_{vn}(\lambda)) \nonumber\\
    =\ & i\partial_x\left(-\nabla_{\textbf{k}}\theta\cdot\frac{\partial\textbf{k}_{vn}}{\partial\lambda}\right) + i\frac{\partial\textbf{k}_{vn}}{\partial\lambda}\cdot\nabla_{\textbf{k}}(\partial_x\theta)\nonumber\\
    =\ & -i\frac{\partial y_{vn}}{\partial\lambda}(\partial_x\partial_y - \partial_y\partial_x)\theta(\textbf{k}-\textbf{k}_{vn}(\lambda)) \nonumber \\
    =\ & -2\pi iC\delta_P(\textbf{k}-\textbf{k}_{vn}(\lambda))\frac{\partial y_{vn}}{\partial\lambda},
    \label{dxdl-dldx}
\end{align}
where $\partial_x$, $\partial_y$ are taken with respect to  $k_x$ and $k_y$ respectively, and $y_{vn}$ is the $y$-component of $\textbf{k}_{vn}$. In the first step, we have used the fact that the derivatives commute except when applied to the phase.
Note that we used (\ref{curlgrad}) for the third equality. A similar expression can be found for the other component. Then substituting the expression found for the adiabatic current density back into 
(\ref{poldefapp}) we obtain the desired result for the polarization (\ref{chernpol}).

\bibliography{biblio}
\end{document}